# Adsorption property of fatty acid on iron surface with Σ3(111) grain boundary

Ivan Lobzenko[1], *Yoshinori Shiihara[1], Yoshitaka Umeno[2], Yoshikazu Todaka[3]

1. Toyota Technological Institute (166 Hisakata, Tempaku, Nagoya, 468-0034, Japan)

*E-mail: shiihara@toyota-ti.ac.jp

2. Institute of Industrial Science, The University of Tokyo
(Komaba, Meguro-ku, Tokyo 153-8505, Japan)
3. Department of Mechanical Engineering, Toyohashi University of Technology,
(Toyohashi Aichi 441-8580, Japan)

Reducing the coefficient of boundary friction on steel surfaces is one of key technologies to improve the efficiency of machines such as automotive engines. It has been shown that the boundary friction on nanostructured steel surfaces in the sliding test using hydrocarbon lubricant molecules is smaller than the friction of normal steel surfaces. The main difference between the nanostructured and a normal surfaces is the density of grain boundaries and other surface defects. The surface defect can attract lubricant molecules and enhance lubricating film formation on metal surfaces. This can be one of the mechanisms that induce the friction reduction on the nanostructured steel surface. In this work, using first-principles calculations, the adsorbability of a lubricant molecule, a fatty acid, on a defected iron surfaces has been studied. Adsorbability of the Fe(110) surface with symmetrical tilt Σ3(111) grain boundary was compared to that of the clean Fe(110) surface. As a result, we found that the molecule is adsorbed on sites close to the grain boundary more strongly (0.77 eV in average) than on the Fe surface without grain boundary.

- Introduction

Iron is an indispensable for industry due to its low cost and high strength. Chemical properties of iron surfaces are decisive for production and use of durable steel. As a transition metal, iron is active chemical agent [1,2]. Currently, adsorption on iron surfaces are being actively studied theoretically with the use of quantum-mechanical modeling, which provides better understanding of such processes as oxidation [3,4], carburization [5], etc. [6,7,8]. Due to the advances in recent hardware and computational programs, it became possible to model a molecule interactions with iron surface [9,10,11], which is a step towards the understanding of interactions between an iron surface and a lubricant.

The coefficient of boundary friction on steel surfaces is in direct dependency on how an iron surface is reacting with the used lubricant. It is well known that the average grain size is affecting wide range of steel properties, and unique characteristics are associated with the smaller grain size. Steels with ultra small grains are also called hetero-nanostructured or ultrafine-grained steels, and can be obtained by applying a severe plastic deformation to the ordinary alloy [12]. Surface reactivity of such materials is largely affected by the high concentration of grain boundaries (GB) and other crystal defects in the vicinity of GBs. Tribological properties of hetero-nanostructured steels are being extensively studied in recent years [13-18]. It has been experimentally shown that the boundary friction on hetero-nanostructured steel surfaces is smaller than the friction of normal steel surfaces [17]. Yet such an effect has not been explained theoretically.

The authors have already confirmed an important role of the GB in adsorbability of single oxygen and carbon atoms on the iron surface [19]. In the present paper the adsorption of acetic acid ion (acetate) molecule on the Fe surface with and without the GB is studied by means of first principal simulations.

This paper is organized as follows. First we present the model and the method of studies. Then main results are discussed in order to draw out the difference between the surface with and without the grain boundary.

- Simulation Model

Two types of body centered cubic (bcc) Fe surfaces are discussed. The first one is a defect free (clean) Fe (110) surface. The slab model consists of 63 Fe atoms arranged in 7 atomic layers stacked in the $z$ direction, the $xy$ projection of the slab can be seen in Fig. 1 (a). The computational cell also contains 15 Å of the vacuum, which is needed to avoid the self-interaction. The second type of the surface is the Fe (110) surface with the $\Sigma 3(111)$ grain boundary representing the nanostructured surface (Fig. 1 (b)). The slab contains 192 iron atoms arranged in 4 layers and 15 Å vacuum region stacked in the $z$ direction. Due to the periodic boundary conditions, the computational cell has two grain boundaries, so that it represents an infinite crystal in the $xy$ plane.

The acetic acid, which is the simplest fatty acid, is chosen as the model adsorbate molecule. Fatty acids adsorb on the metal via chemical bonding [20,21]. To model the bonding between the adsorbate and the metallic surfaces, acetic acid ion (acetate), obtained by removing one hydrogen atom, was used as shown in Fig. 1 (c).

Five adsorption sites are set on the clean surface. The names of sites are conventional and define the initial position of the carbon atom C1 and the orientation of the O1-C1-O2 surface; atoms notation is given in Fig. 1 (c). Five adsorption sites are also set on the surface with the GB. Sites 1, 2, and 3 are unique and appear only in the system with the GB, while sites 4 and 5 resemble the middle bridge position on the (110) surface (see Fig. 1b). However, the sites 4 and 5 locate close to the GB and therefore are showing how the vicinity of the grain boundary affects the adsorption.

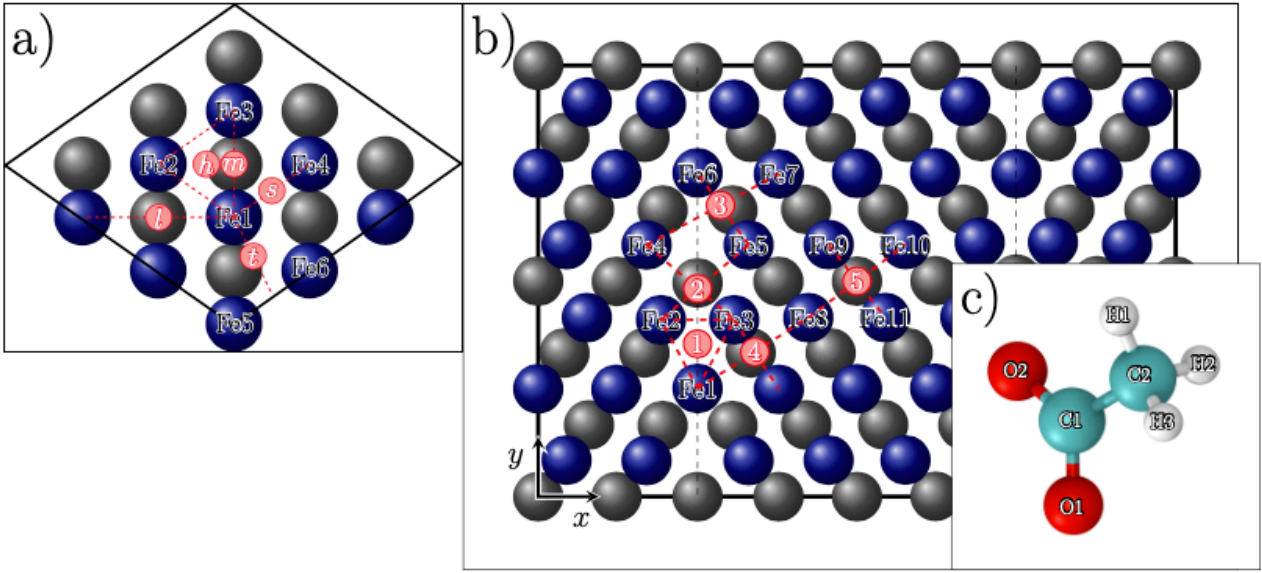

Fig. 1 *(Color online)* Top view (*xy* projection) of (a) Fe(110) surface with the Σ3(111) grain boundary and (b) Fe(110) clean surface. Considered adsorption states are shown by red circles; names of sites for the clean surface are: *h* – hollow, *l* – long bridge, *m* – middle bridge, *s* – short bridge, *t* – top to bridge. Gray circles correspond to the second atomic layer. Figure (c) shows the structure of the acetate molecule. Atoms numbers are used later in the text.

- Calculation Detail

Simulations were performed using a quantum mechanical approach. The density functional method based on the linear combination of atomic orbital (LCAO) method was used, as implemented in the Open source package for Material eXplorer (OpenMX) [22]. The exchange-correlation energy was taken in the form of the generalized gradient approximation (GGA) [23]. The energy cutoff, which controls the accuracy through the numerical integrations and the density of the fast Fourier transform grid [24], was set to 400 Ry. Also, norm-conserving pseudopotentials were used [25]. In the energy minimization process all structures were optimized until the forces on each atom were less than 0.01 eV/Å. The Brillouin zone integration for the clean surface was performed using 5×5×1 Monkhorst-Pack *k*-points mesh [26]. The Brillouin zone integration for the surface with the grain boundary was performed using only the Γ point in the reciprocal space. The adsorption energy $E_{\text{ads}}$ for each surface was calculated as

$$E_{\text{ads}} = E_{\text{surf+mol}} - E_{\text{surf}} - E_{\text{mol}} ,$$

where $E_{\text{surf+mol}}$ is the energy of the system with the molecule attached to the iron surface while $E_{\text{surf}}$ and $E_{\text{mol}}$ are energies of the Fe surface and single molecule in vacuum, respectively.

During the process of atomic positions optimization, the structure converges to the local minimum. For each adsorption site, we used several initial orientations of the molecule randomly placed close to the perfect initial adsorption position in order to obtain the lowest energy after the optimization and associate it with the adsorption site.

The employed LCAO method opens possibility to decompose the total energy into atomic contributions. Due to the fact that each basis function can be attributed to a particular atom, the projections of the total energy onto the basis functions could be sorted with respect to atoms. OpenMX package is able to produce such decomposition.

- Results and discussion

Optimized geometry of the acetate molecule at the particular adsorption site can be described with the help of the interatomic distances and angles, as shown in Fig. 2. Most valuable data is lengths of the C-C and C-O bonds. Among the angles, we highlight the inclination $\theta$ and azimuth $\varphi$ of the C1–C2 bond and the angle between that bond and O1–C1–O2 plane ($\alpha$). Those parameters for all adsorption sites on the clean surface and the surface with the GB are given in Table 1 and Table 2, respectively.

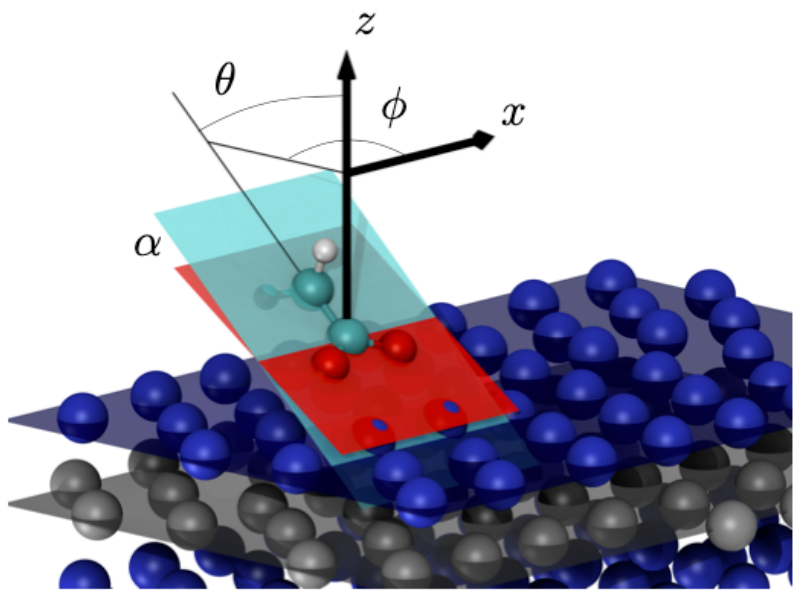

Fig. 2 *(Color online)* Acetate molecule on Fe(110) surface with the Σ3(111) grain boundary with inclination $\theta$ and azimuth $\varphi$ of the C1–C2 bond. Carbon atoms belong to the plane colored cyan. The plane colored red contains two oxygen atoms and one carbon (O1–C1–O2 plane). The angle $\alpha$ is acute angle between the two planes.

After the structural optimization the molecule is placed differently on different adsorption sites. Conformations for the clean Fe (110) surface are shown in Fig. 3. It is worth mentioning that in the case of the "hollow" site on the clean Fe (110) surface only one oxygen atom is forming the bond with the substrate (see Fig. 3a), which is reflected in the difference of the carbon – oxygen bonds (C1 – O1 and C1 – O2) lengths, while for all other adsorption sites bonds lengths are equal. The parameter $\alpha$ shows how much the molecule is bowed: that parameter is raising with the increase of the inclination angle ($\theta$), which shows how much the molecule deviates from the $z$ axis.

The adsorption energies for the molecule on the clean surface and the surface with the GB are shown in Table 3 and Table 4. Correspondent structures are shown in Fig. 4 The main result is that all adsorption energies on the surface with the grain boundary are lower than any of the energies on the clean surface. The averaged difference is 0.77 eV, which shows that the presence of the grain boundary enhances adsorption properties significantly.

Table. 1 Geometry of the acetate on clean Fe(110) surface.
Distances are given in [Å], angles are in [degrees].

|  | hollow | long bridge | middle bridge | short bridge | top to bridge |
|---|---|---|---|---|---|
| C1 - C2 | 1.50 | 1.50 | 1.51 | 1.51 | 1.50 |
| C1 - O1 | 1.38 | 1.29 | 1.28 | 1.28 | 1.28 |
| C1 - O2 | 1.21 | 1.29 | 1.28 | 1.28 | 1.28 |
| $\theta$ | 64 | 38 | 1 | 1 | 20 |
| $\phi$ | 83 | 129 | -127 | -60 | -129 |
| $\alpha$ | 0 | 2 | 0 | 0 | 3 |

Table. 2 Geometry of the acetate on Fe(110) surface with Σ3(111) grain boundary.
Distances are given in [Å], angles are in [degrees].

|  | Site 1 | Site 2 | Site 3 | Site 4 | Site 5 |
|---|---|---|---|---|---|
| C1 - C2 | 1.51 | 1.50 | 1.51 | 1.50 | 1.51 |
| C1 - O1 | 1.28 | 1.30 | 1.28 | 1.27 | 1.28 |
| C1 - O2 | 1.28 | 1.30 | 1.28 | 1.28 | 1.28 |
| $\theta$ | 7 | 48 | 19 | 2 | 1 |
| $\phi$ | 91 | 90 | 161 | 12 | -93 |
| $\alpha$ | 1 | 7 | 3 | 2 | 1 |

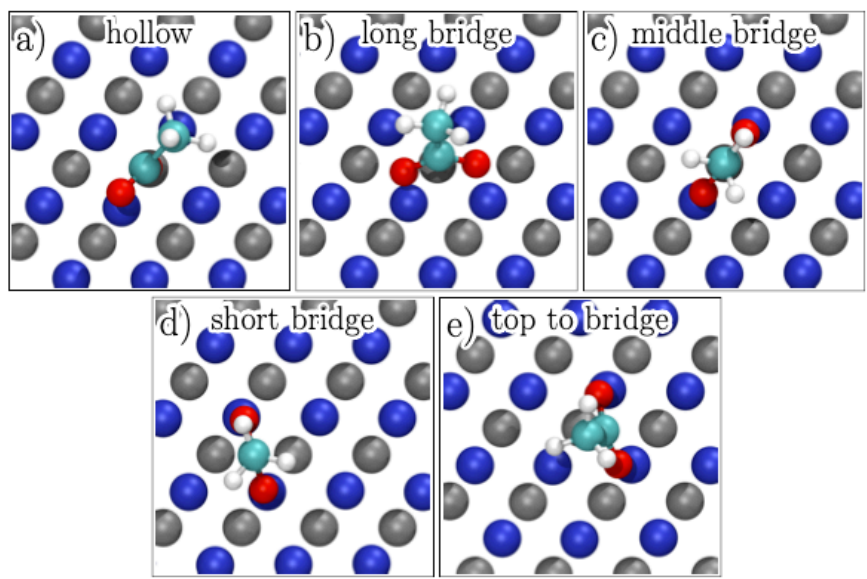

Fig. 3 *(Color online)* Top view (*xy* projection) of the acetate molecule on adsorption sites of the clean Fe(110) surface. Atoms of the top layer of the slab are shown in blue, while gray circles denote atoms of the second layer

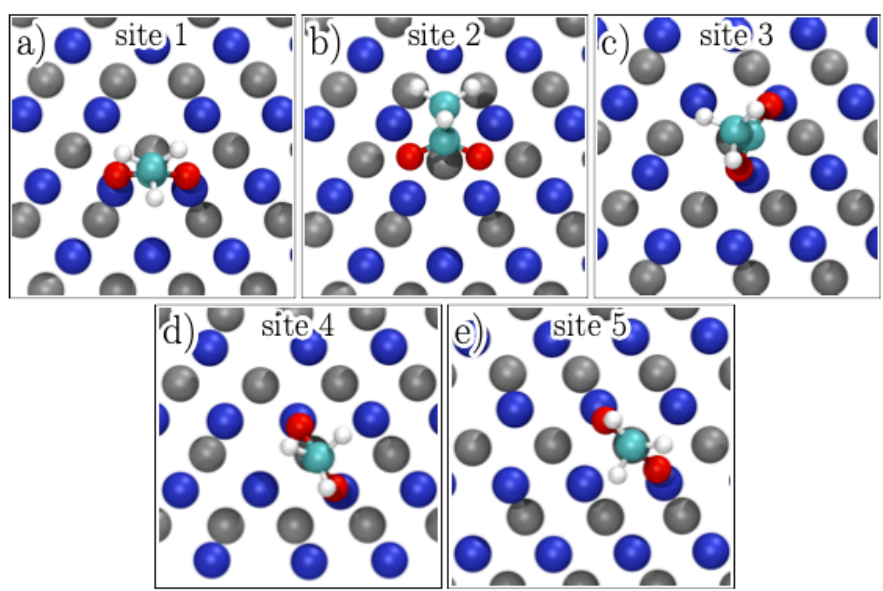

Fig. 4 *(Color online)* Top view (*xy* projection) of the acetate molecule on adsorption sites of Fe surface with the grain boundary. Atoms of the top layer of the slab are shown in blue, while gray circles denote atoms of the second layer

As it has been already said, the approach based on orbital energy decomposition allows us to quantitatively estimate the energy contribution from each individual atoms as shown in Fig. 5. We collected contribution from each atomic layer of the slab and from all atoms of the same type for the molecule (see Table 3 and Table 4). That data can be used to analyze the difference in the energies brought by the slab and the molecule. It is seen that the contribution to the adsorption energy of the molecule is negative, while the contribution of the slab is positive, also for sites on the surface with the GB all energies are lower than energies for the clean surface. In other words, on the surface with the GB after the structural optimization the molecule is able to find more preferable state, while the slab is less disturbed. Comparing site 2 in Fig 5 (a) and the long bridge site on the "clean surface in Fig. 5 (b), one can see that the carbon atom C1 is always the one with the biggest negative energy difference.

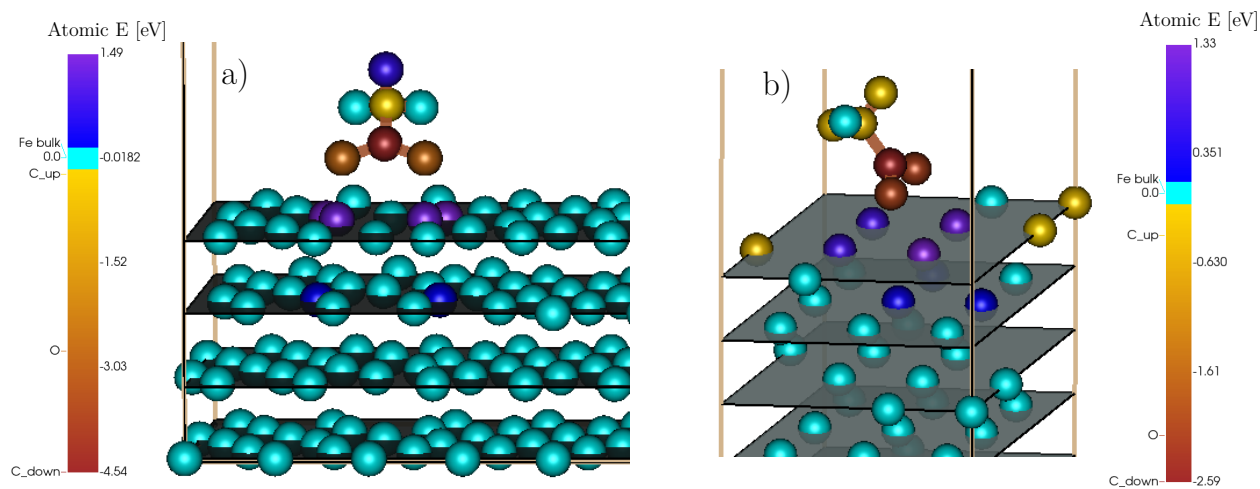

Fig. 5 *(Color online)* Atomic contributions to the adsorption energy for (a) the site 2 on the iron surface with grain boundary and (b) the long bridge site on the "clean" iron surface. Atoms denoted C_up, C_down correspond to atoms C2, C1 in Fig. 1(c).

Table 3 Adsorption energies (in [eV]) on clean Fe(110) surface.

| Contributions | hollow | long bridge | middle bridge | short bridge | top to bridge |
|---|---|---|---|---|---|
| layer 1 up | 3.11 | 3.35 | 2.17 | 2.14 | 2.41 |
| layer 2 | 0.70 | 0.84 | 0.62 | 0.60 | 0.59 |
| layer 3 | 0.11 | 0.14 | 0.16 | 0.15 | 0.12 |
| layer 4 | -0.01 | -0.02 | -0.01 | -0.02 | -0.09 |
| layer 5 | -0.36 | -0.36 | -0.35 | -0.35 | -0.34 |
| layer 6 | 0.08 | 0.07 | 0.07 | 0.07 | 0.33 |
| layer 7 low | 0.17 | 0.08 | 0.08 | 0.08 | 0.02 |
| all slab Fe | 3.79 | 4.10 | 2.73 | 2.67 | 3.04 |
| atoms O | -2.54 | -4.31 | -2.99 | -3.00 | -3.28 |
| atoms C | -4.07 | -2.97 | -3.18 | -3.06 | -2.83 |
| atoms H | -0.66 | -0.51 | -0.23 | -0.25 | -0.37 |
| all molecule | -7.26 | -7.79 | -6.41 | -6.32 | -6.48 |
| E adsorption | -3.46 | -3.69 | -3.68 | -3.65 | -3.44 |

Table 4 Adsorption energies (in [eV]) on Fe(110) surface with Σ3(111) grain boundary.

| Contributions | Site 1 | Site 2 | Site 3 | Site 4 | Site 5 |
|---|---|---|---|---|---|
| layer 1 up | 2.52 | 4.58 | 2.69 | 2.29 | 2.48 |
| layer 2 | 0.33 | 0.65 | 0.58 | 0.51 | 0.77 |
| layer 3 | 0.21 | -0.11 | -0.03 | 0.03 | -0.04 |
| layer 4 low | -0.09 | 0.02 | 0.01 | 0.01 | -0.03 |
| all slab Fe | 2.96 | 5.14 | 3.24 | 2.84 | 3.18 |
| atoms O | -3.20 | -5.59 | -3.16 | -2.87 | -3.01 |
| atoms C | -5.18 | -4.78 | -5.18 | -5.22 | -5.43 |
| atoms H | 1.02 | 0.59 | 0.86 | 1.01 | 1.02 |
| all molecule | -7.36 | -9.78 | -7.48 | -7.09 | -7.42 |
| E adsorption | -4.40 | -4.64 | -4.24 | -4.25 | -4.24 |

To discuss the electron clouds rearrangement induced by the adsorption we present the 2D map of the charge density difference ($\Delta\rho = \rho_{surf+mol} - \rho_{surf} - \rho_{mol}$) for the most preferable adsorption site 2 on the Fe (110) with GB. Two sections are made. Fig. 6 (a,b) shows the section by the plane in which the O – Fe chemical bonds lay (atoms Fe2, Fe3, and O1, O2 denoted in Fig. 1 form that plane). The rearrangement of the electronic clouds happens in the way that negative charge is concentrated on top of the iron surface, while oxygen atoms of the molecule gain positive charge shaped towards the surface. Therefore, the essential character of the bonding is ionic. The second section, shown in Fig. 6 (c,d), is perpendicular to the first one. It is made using the plane shared by atoms C1, C2 and Fe6. The tilt of the molecule towards the iron surface is clearly observed on that section. The main bonds between oxygen and iron atoms are not shown, however the rearrangement of the electronic clouds around atoms C1 and C2 is better represented. That rearrangement results in the appearance of two additional bonds: between atom C1 and atom Fe6, and between atom C1 and atom $Fe_{II}1$ (the atom from the second layer). The additional interaction further decreases the energy of C1 atom (Fig. 5(a)). As a result, electronic clouds are affected more in adsorption on the surface with grain boundary than in adsorption on the clean Fe (110). The molecule appears to form stronger bonds. That is why the energy contribution of the molecule increases in absolute value, and, being negative, results in lower adsorption energy.

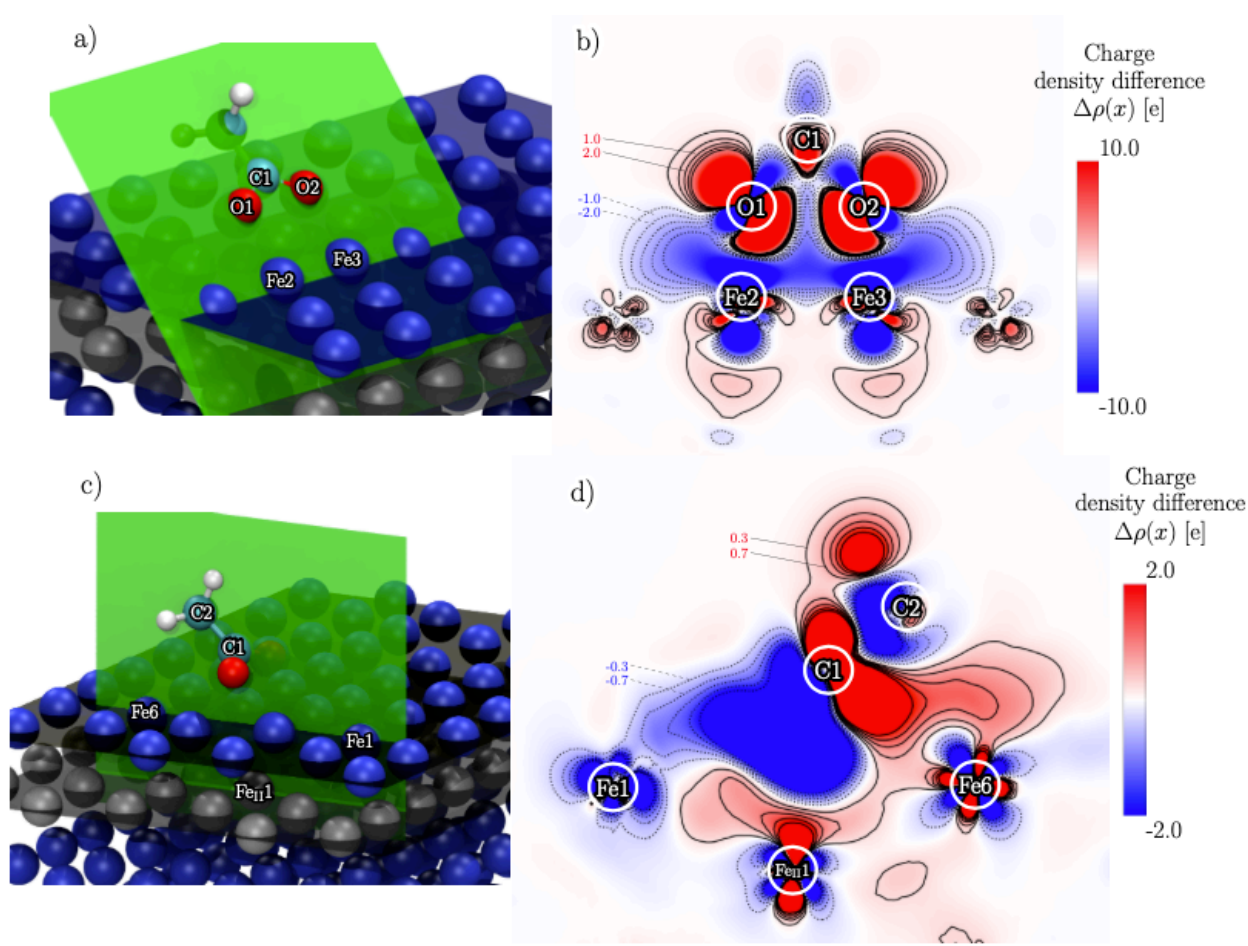

Fig. 6 *(Color online)* Charge difference ($\Delta\rho = \rho_{surf+mol} - \rho_{surf} - \rho_{mol}$) map for acetate on site 2 on Fe (110) surface with $\Sigma3(111)$ grain boundaries. Section containing atoms O1, O2, Fe2, Fe3 is schematically shown in (a), while the charge difference map is shown in (b). Section containing atoms C1, C2, Fe1, Fe6 and $Fe_{II}1$ is shown in (c), while the charge difference map is shown in (d). Atoms notation coincides with the notation in Fig. 1

It is also noticeable that higher order of rearrangement of electronic clouds around the atom C2 puts the hydrogen atoms of the molecule in a less preferable state, as it is seen from the fact that on the surface with GB hydrogen atoms contribute positive energy to the adsorption (see Fig. 5 (a)). However the total gain from the additional bonds between the carbon atoms and the iron surface exceeds that, and ultimately the molecule state appears to be more preferable on the sites near the GB.

- Conclusions

Adsorption of the acetic acid ion on the Fe(110) surface with Σ3(111) grain boundary has been studied showing enhanced properties compare to the correspondent Fe(110) "clean" iron surface. The adsorption energy gain is higher for adsorption sites located closer to the GB, however the effect of enhancement of the adsorption was confirmed even at a relatively distant position (approximately 5 Å). The energy contributions of atoms of different type is shown to be qualitatively different for the adsorption on the surface with GB, which is connected to the higher order of the electronic clouds rearrangement. The increase in chemical activity due to the presence of the GB can be clearly concluded.

Due to high demand on the calculation resources, we considered only relatively small structures. It is particularly needed to investigate the same effect for systems with bigger grains, different grain boundaries, and with more than one molecule as an adsorbate. That will be the subject of further studies.